\PassOptionsToPackage{authoryear,round}{natbib}
\documentclass[article,onecolumn,english,floatfix,longbibliography]{revtex4-1}
\usepackage{blindtext}
\usepackage{graphicx}
\usepackage[utf8]{inputenc}

\usepackage{epstopdf, epsfig}
\usepackage{amsmath}
\usepackage{float}
\usepackage{stmaryrd}
\usepackage{braket}
\usepackage{xcolor}
\usepackage{subcaption}
\usepackage{comment}
\usepackage[switch]{lineno} 

\makeatletter
\renewcommand{\citet}[1]{%
  \def\@citea{}
  \@for\@citeb:=#1\do{%
    \@citea\def\@citea{; }
    \citeauthor{\@citeb} (\citeyear{\@citeb})%
  }%
}

\makeatletter
\newcommand{\@citecomma}{, } 
\renewcommand{\citep}[1]{%
  \begingroup
  \let\@citea\@empty
  \textup{(}%
  \@for\@citeb:=#1\do{%
    \@citea
    \citeauthor{\@citeb}, \citeyear{\@citeb}%
    \let\@citea\@citecomma
  }%
  \textup{)}%
  \endgroup
}
\makeatother

\usepackage{bm}
\newcommand{\ds}{\displaystyle}

\renewcommand{\Delta}{\triangle}

\newcommand {\tck}{\textcolor{black}}

\newcommand {\tcw}{\textcolor{white}}
\newcommand {\mb}{\boldsymbol}

\begin{document}
\title{New constraint on Europa's ice shell: magnetic signature from the ocean}
\author{F.Daniel $^{1,2}$}
\email{florentin.daniel@northwestern.edu}
\author{L.Petitdemange$^3$}
\email{ludovic.petitdemange@upmc.fr}
\author{C. Gissinger$^{1,4}$}
\email{christophe.gissinger@phys.ens.fr}

\affiliation{
$^1$ Laboratoire de Physique de l'Ecole normale supérieure (LPENS), ENS, Universite PSL, CNRS, Sorbonne Université, Université de Paris, Paris, France\\
$^2$ Center for Interdisciplinary Exploration and Research in Astrophysics (CIERA), Northwestern University, Evanston, IL, USA\\
$^3$ Laboratoire d'Instrumentation et de Recherche en Astrophysique (LIRA), Observatoire de Paris, Université PSL, Sorbonne Université, Université Paris Cité, CY Cergy Paris Université, CNRS,75014 Paris, France \\
$^4$ Institut Universitaire de France (IUF), Paris, France}

\begin{abstract}

Jupiter’s icy moons are believed to host subsurface liquid oceans, and among them, Europa stands out as one of the most promising candidates for extraterrestrial life. Yet, the processes driving oceanic flows beneath its ice shell, as well as the factors controlling the thickness of this ice, remain incompletely understood. One especially distinctive feature of Europa is that its salty ocean is electrically conducting and thus influenced by Jupiter’s time-varying magnetic field, which is believed to drive a large-scale zonal flow. Here, we examine how this magnetically-induced jet affects both the heat flux and the dynamics of the convective flow  within Europa’s ocean.
We first show that the magnetically-driven jet efficiently transports heat in stably stratified regions near the top of the ocean, and may alter the expected convective scaling laws in deeper layers. Second, by analysing the latitudinal distribution of heat flux and relating it to ice-thickness variations, we make predictions that can be compared with current observations. In anticipation of the upcoming JUICE and Europa Clipper missions, we discuss how improved measurement precision could help further constrain the ocean's properties and refine our model-based forecasts.
\end{abstract}

\maketitle

\section{Introduction}
\tck{Past observations suggest that Europa is composed of a superficial ice layer lying on the top of a salty liquid ocean \citep{Pappalardo1999, Khurana1998}. Current estimates for its internal structure suggest the presence of a metallic core of about $\ds 500\,km$ just below a rocky mantle whose thickness is of the order of $\ds 1000\,km$. Many different sources of heat flux acting on the water have been considered (see next), leading to an ocean whose size varies between $\ds 80$ and $\ds 170\,km$ (see for instance \citet{Soderlund2020}), and $\ds 15$ to $\ds 50\,km$ for the ice. \tck{Horizontal} variations, i.e. both in longitudes and latitudes can also be expected, of the order of $\ds 10\,km$ in the deeper ice scenario. Several unknowns indeed remain concerning the precise geometry as well as the structure of the ocean and ice layers. \\
\tck{Via precise gravity measurements, ice-penetrating radars or magnetometers,} future space missions JUICE \citep{Grasset2013} and Europa Clipper \citep{Phillips2014} therefore offer an exciting prospect as they will provide the community with data measurements on the structure of \tck{the superficial layer, through for instance a characterisation of the ice thickness}. With the latter lying on top of the liquid water, its knowledge and understanding can be used as an indirect way to infer some of the ocean's properties. Ocean currents can influence subsurface exchange processes and the melting of the ice \citep{Ojakangas1989, Nimmo2007}. A realistic model of Europa's ocean is thus of the utmost importance to properly assess its ability to affect the ice through the flux coming at the interface between the two zones.} \\ 

\tck{Icy-moon oceans share many characteristics with classical geophysical fluid dynamics problems, as they involve rotating, saline water contained within spherical shells, yet they also exhibit notable differences with Earth’s oceans. Consequently, numerous studies have examined these extraterrestrial oceans, incorporating various physical processes that may drive ocean currents. While thermal convection is often considered the most natural source of oceanic motion, other mechanisms, such as inertial waves \citep{Rovira2019} and libration \citep{Rekier2019}, have also been proposed. Thermal fluxes originating from the rocky mantle - ranging from approximately $\ds 6$ \citep{Tobie2003} to $\ds 46\, mW/m^2$ \citep{Howell2021}, depending on the balance of radiogenic and tidal heating - are generally regarded as the primary driver in thermal convection models \citep{Soderlund2014, Lemasquerier2023, Cabanes2024}.}
The many uncertainties concerning the ocean's properties make it possible that the nature of the buoyancy could however change close to the ice, as the thermal expansion coefficient could become negative due to the low pressure and temperature. This would lead to a layer of stably stratified water to lie on top of a convectively unstable one \citep{Melosh2004}. \tck{The presence of dissolved salts further complicates this stratification, potentially giving rise to double-diffusive phenomena \citep{Vance2005, Wong2022}. More recently, the coupling between ice melting dynamics and ocean circulation obtained through thermal convection has also been investigated in icy-moon contexts \citep{Gastine2025}. Finally, Europa’s proximity to Jupiter subjects it to strong tidal forces \citep{Tyler2008} and magnetic-field interactions \citep{Gissinger2019} - both significant sources of energy dissipation and fluid motions - which further underscore how differently these subsurface oceans behave compared to those on Earth.}

\tck{In this work, we investigate the interplay between thermal convection in Europa’s ocean and the electromagnetic pumping induced by Jupiter’s magnetic field. Accounting for these two effects simultaneously allows us to describe their non-linear couplings, marking a necessary step towards more realistic numerical models.} Through a systematic parametric study, we show how the equatorial zonal flow generated by the Jovian magnetic field in the electrically-conducting ocean \citep{Gissinger2019} modifies properties of rotating convection. \tck{In particular, we identify a new regime in which the magnetically-driven jet significantly alters the latitudinal heat-flux profile, in contrast with purely hydrodynamical studies.} Using theoretical models linking the ice thickness evolution with the flux coming from the ocean, we relate this heat flux contrast to spatial variations of the ice layer, compare our results with current topography measurements, and discuss their implications for future observations. Our numerical model of the ocean is described in \ref{section_method}. Results first focus on Direct Numerical Simulations (DNS), presented in \ref{section_res1}, before presenting the ice thickness profiles obtained in \ref{section_res2}.

\section{Method}
\label{section_method}

\paragraph{Numerical model}
The numerical setup is similar to the one described in \citet{Gissinger2019}; the problem consists in describing the evolution of a layer of liquid water contained between two spheres of radius $\ds r_i$ and $\ds r_o$, simulating the subsurface ocean of Europa which is enclosed between a silicate mantle ($\ds r=r_i$) and a layer of ice ($\ds r=r_o$). The aspect ratio $\ds \chi = r_i/r_o$ is set to $\ds 0.9$. For an outer radius of the ocean $\ds r_o=1550\,km$, it corresponds to a thickness of the liquid layer $\ds d=r_o-r_i=155\,km$ \citep{Soderlund2014}. The reference frame is taken to be the outer sphere which rotates around the $\ds \mb{e_z}$ axis at an angular velocity $\ds \Omega_o$. Although the dynamic of the ice is not considered, its effect on the ocean is modelled by imposing a constant temperature $\ds T_o$ at the outer boundary, set to be that of the ice, uniform over the sphere. On the ocean floor, in order to model the radiogenic flux coming from the disintegration of the silicate, a constant temperature flux $\ds \Phi_i$ is imposed, allowing for thermal stratification to act on the water. \tck{This is achieved using the Boussinesq approximation, similarly to earlier studies (see \citet{Soderlund2014} for a discussion about the relevance to model Europa's ocean using the Boussinesq formalism).} \tck{The fluid is also electrically conducting and is thus influenced by Jupiter’s magnetic field $\ds \mb{B_0}$. Because Jupiter’s magnetic dipole is tilted by about $\ds 10 ^{\circ}$ relative to its rotation axis, Europa experiences a time-varying magnetic field rotating in the azimuthal direction.} It is curl-free and \tck{can} be expressed analytically (see \citet{Gissinger2019} \tck{or Appendix \ref{appendix_B0}}). \tck{The time variation of $\ds \mb{B_0}$ induces electrical currents (or equivalently, a magnetic field perturbation $\ds \mb{b}$) in the moon's salty ocean, thereby generating a Lorentz force that drives motions within the ocean.} In other terms, Europa’s ocean behaves exactly like a gigantic natural electromagnetic pump. The boundary conditions for the velocity are no-slip on both spheres and insulating for the total field $\ds \mb{B_0}+\mb{b}$. \tck{In particular, the use of no-slip conditions implies that we neglect the mechanical coupling between the ice shell and the ocean. However, as explained in \citet{Gissinger2019}, such a coupling can lead to a net torque applied to the ice shell \citep{Hay2023}, notably due to the large-scale magnetically-driven jet. Although this torque could significantly affect the global dynamics, we chose to ignore it here in order to focus on the interaction between thermal convection and magnetic effects within the ocean. In a similar context applied to Saturn’s moon Titan, \citet{Kvorka2022} have shown that the choice between no-slip and stress-free conditions can lead to markedly different results.} \tck{Due to its very low conductivity compared to seawater (typically $\ds 10^{-6}$ to $\ds 10^{-5}\, \mathrm{S/m}$ against $\ds 1\,\mathrm{S/m}$), the ice shell can be accurately modelled as an electrical insulator for the ocean \citep{Schilling2007,Gissinger2019}. Note that the complex rheology of the ice may favour the trapping of salts \citep{Durham2001}, although the ice remains largely insulating relative to ocean water.} \tck{Besides, sharp gradients of magnetic diffusivity near the ice shell may generate magnetic structures independently of flow-induced induction; this effect, though potentially important, is beyond the scope of this study and will be addressed in future work.} 

The magnetohydrodynamics (MHD) equations describing the evolution of the velocity $\ds \mb{u}$, modified pressure $\ds \Pi$, temperature perturbations $\ds \Theta$ and induced magnetic field $\ds \mb{b}$ of the flow in the Boussinesq approximation are then :

\begin{align}
	\nonumber \frac{\partial \mb{{u}}}{\partial {t}} + \mb{{u}} \cdot {\nabla} \mb{{u}}  + 2 \mb{e_z}\times \mb{{u}} &= - {\nabla} \Pi + Ek {\Delta} \mb{{u}} + \frac{Ra^Q Ek^2}{Pr}\Theta /r^2 \mb{e_r} \\
	&+ \widetilde{\Lambda} Ek\left({\nabla} \times \mb{{b}}\right) \times (\mb{B_0}+\mb{b}), \label{MTH_eq_U_adim1}\\
	\frac{\partial {\Theta}}{\partial {t}} +  \mb{{u}} \cdot {\nabla} {\Theta} &= \frac{Ek}{Pr} {\Delta} {\Theta} +\frac{u_r}{r^2}, \label{MTH_eq_THETA_adim1}\\
	\frac{\partial \mb{{b}}}{\partial {t}} &= {\nabla} \times \left( \mb{{u}}\times (\mb{B_0}+\mb{b}) \right) + Ek_\eta {\Delta} \mb{{b}} - \frac{\partial \mb{B_0}}{\partial t}, \label{MTH_eq_B_adim1}\\
	{\nabla} \cdot \mb{{u}} &= 0, \label{MTH_eq_DIVU_adim1}\\
	{\nabla} \cdot \mb{{b}} &=	{\nabla} \cdot \mb{{B_0}}= 0, \label{MTH_eq_DIVB_adim1}
\end{align}

and have been made without dimension introducing the following dimensionless parameters:

\begin{equation}
	\widetilde{\Lambda} = \frac{B_0^2}{\rho \mu_0 \nu \Omega_o},\, Ek = \frac{\nu}{r_o^2\Omega_o},\, Ra^Q = \frac{\alpha_T g_o \Phi_i r_o^2r_i^2}{\nu \kappa k}, \,Pr = \frac{\nu}{\kappa}, \,Ek_\eta = \frac{\eta}{r_o^2\Omega_o},
	\label{MTH_eq_nb_sans_dim1}
\end{equation}

respectively the \tck{modified} Elsasser, Ekman, Rayleigh, Prandtl and magnetic Ekman numbers. In their definition appear some properties of the ocean, such as its density $\ds \rho$ and kinematic ($\ds \nu$), thermal ($\ds \kappa$) or magnetic ($\ds \eta$) diffusivities. $\ds k$ is the thermal conductivity, $\ds \alpha_T$ the thermal expansion coefficient and $\ds g_o$ the value of gravity at the boundary between the ocean and the ice, \tck{whose profile is $\ds g(r)=g_o(r_o/r)^2$. The background temperature profile, implicitly used in (\ref{MTH_eq_THETA_adim1}) through its gradient, is $\ds 1/r-1$ with our choice of units}. $\ds B_0$ is the typical amplitude of Jupiter's field in the vicinity of Europa, \tck{$\ds \omega = 7.8 \Omega_o$ its pulsation in the frame of the satellite}. All these parameters are considered constant during a simulation. We set $\ds Pr=12$ and $\ds Ek_\eta=10^{-2}$ in all this study (see Table \ref{table_method}). Note that our expression of $\ds Ek$ differs from classical studies of rotating convection where the length used is the height of the ocean $\ds d$. The two expressions are related as $\ds Ek^d=Ek/(1-\chi)^2$. The equations are integrated using the pseudo-spectral code PaRoDy \citep{Dormy1998, Aubert2008}, using the ShtNS library \citep{Shtns2013}.\\
\paragraph{Observables}

We define a Reynolds number $Re_{polo}^{NA} = \sqrt{2E_{kin,polo}^{NA}}r_o/\nu$ based on the non-axisymmetric poloidal kinetic energy averaged over the volume of the fluid $\ds E_{kin,polo}^{NA} = (1/2V)\int_V (u_{polo}^{NA})^2 dV$. \tck{Monitoring this component allows us to isolate the contribution of thermal convection, which primarily drives non-axisymmetric poloidal flows, as opposed to the largely axisymmetric flows induced by magnetic forcing.} The Nusselt number $\ds Nu$ is measured by using the evolution of the difference of temperatures between the two spheres, as we consider convection with an imposed heat flux. Following \citet{Mound2017}, it is defined as $\ds Nu = 1/\left(1+\frac{k}{\Phi_i d \chi}\left(\langle \Theta\rangle_{S(r_i)} -\langle \Theta \rangle_{S(r_o)} \right)\right)$, where $\ds \langle .. \rangle _{S(r)}$ stands for the average on a sphere of radius $\ds r$. \tck{All} observables reported are time averaged once a statistically stationary state of the system is reached. We relate the flux based Rayleigh number to the (classical) one as $\ds Ra= \left(Ra^Q/Nu\right) \left(1-\chi \right)^4/\chi$.
\paragraph{Table}

\begin{table}[H]
\begin{center}
\begin{tabular}{l|l|l}
Parameters & Values (Europa)     & Values (DNS)      \\
\hline 
$Ra^Q$                 & $10^{31}-10^{32}$   & $10^5-10^{12}$    \\
$Ek^d$                 & $10^{-12}-10^{-11}$ & $10^{-3}-10^{-2}$ \\
$\widetilde{\Lambda}$  & $10$                & $1-3.10^3$        \\
$Ek_\eta$              & $10^{-3}-10^{-2}$   & $10^{-2}$         \\
$Pr$                   & $12$                & $12$              \\
$\chi$                 & $0.92-0.94$         & $0.9$       \\
$Ro_c$                 & $10^{-3}-10^{-1}$   & $10^{-3}-10$              
\end{tabular}
\end{center}
\caption{Comparison between the estimated parameters for Europa \cite{Soderlund2019, Gissinger2019, Lemasquerier2023} and their values in our simulations.}
\label{table_method}
\end{table}

The intrinsic limitations of current DNS make it impossible to simultaneously match all relevant physical parameters for Europa’s ocean (see Table \ref{table_method}). \tck{Indeed, the kinematic and thermal diffusivities are the most limiting factors as seen from the values of $\ds Ra^Q$ and $\ds Ek^d$, meaning that our simulations describe poorly the dynamics of their corresponding boundary layers.} Nevertheless, \tck{the bulk dynamics of the ocean can be accurately captured as some} parameters, such as $\ds Ek_\eta$ or $\ds \widetilde{\Lambda}$ match the real values, meaning that our simulations describe well \tck{ohmic diffusion} or the ratio of the Lorentz and viscous forces respectively. The ratio of buoyancy to Coriolis forces $\ds Ro_c = \sqrt{Ra/Pr}Ek^d$ is also accurately reproduced despite Ekman numbers orders of magnitude higher that realistic estimates. On the contrary, the balance of Lorentz to inertia forces $\ds \widetilde{\Lambda}Ek$ is poorly described.
The numerous regimes the ocean hosts can nonetheless be approached by simulating the ocean in appropriate regions of the space parameters and extrapolating to real systems.

\section{Results}
\label{section_results}

\subsection{DNS outputs}
\label{section_res1}

\begin{figure}[H]
    \begin{subfigure}{0.45\textwidth}
        \includegraphics[width=9.0cm]{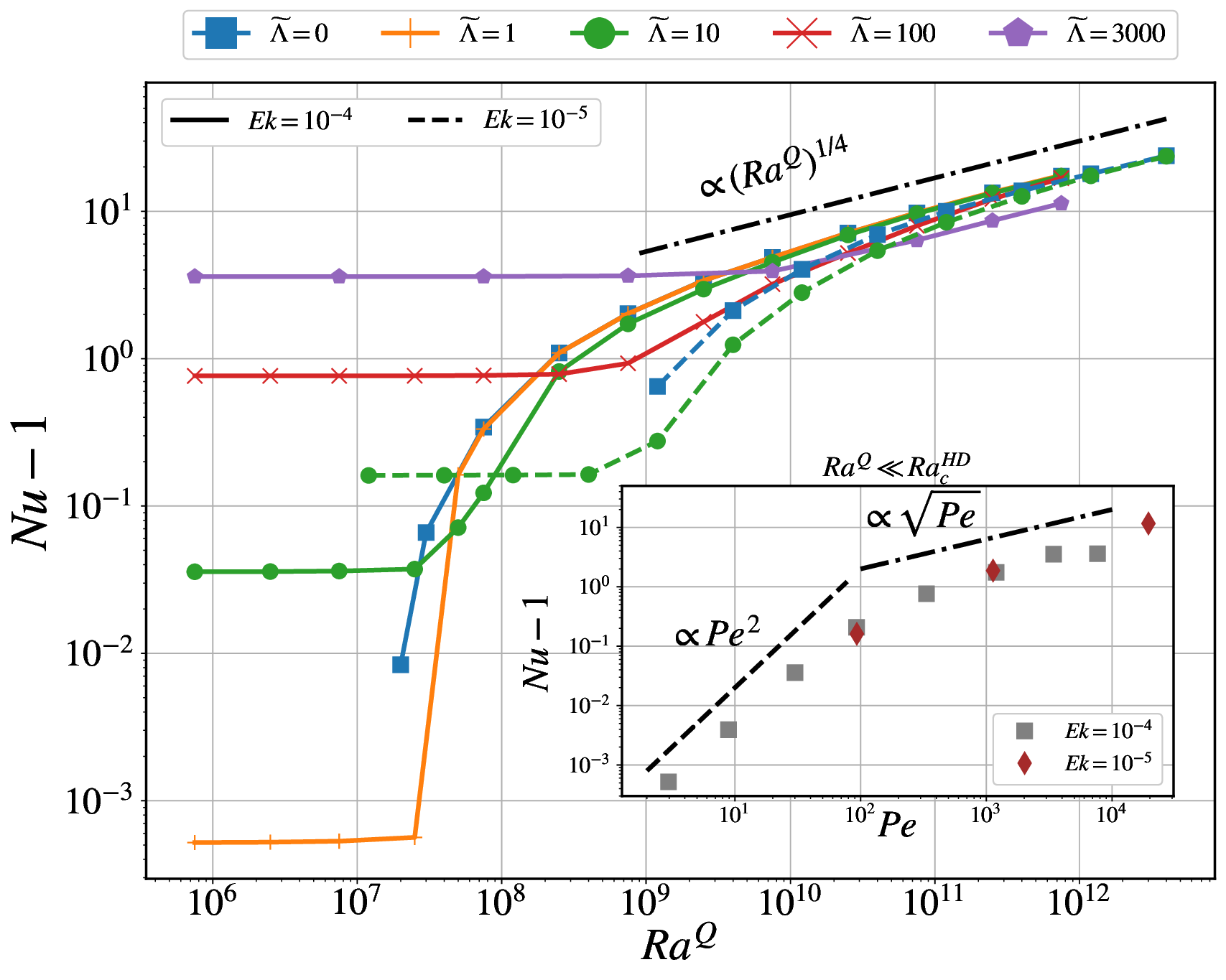}
        \caption{}
        \label{fig1a}
    \end{subfigure} \hspace{0.1\textwidth}
    \begin{subfigure}{0.55\textwidth}
        \includegraphics[width=9.0cm]{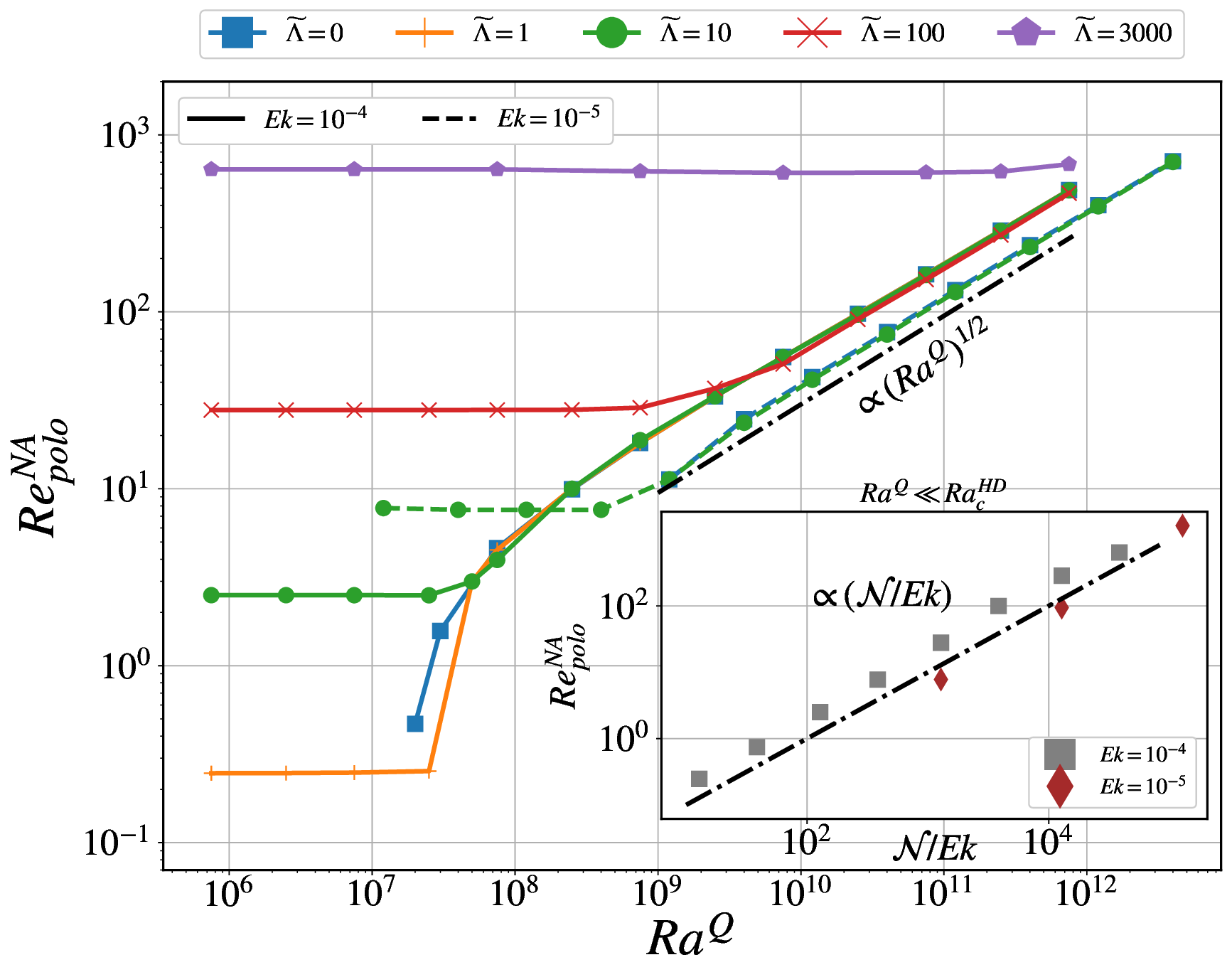}
        \caption{}
        \label{fig1b}
    \end{subfigure}
\caption{Evolutions of $\ds Nu$ (left) and $\ds Re^{NA}_{polo}$ (right) against $\ds Ra^Q$ for $\ds Ek=10^{-4}$ and $\ds Ek=10^{-5}$ with varying values of $\ds \widetilde{\Lambda}$. For large values of $\ds Ra^Q$, all the simulations reported approach the behaviour of the hydrodynamical branch $\ds Nu \propto (Ra^Q)^{1/4}$ and $\ds Re^{NA}_{polo} \propto (Ra^Q)^{1/2}$ respectively. \textit{Insets :} Same quantity as the main Figures, but evolving as a function of $\ds Pe =Pr Re^{NA}_{polo}$ (left), or $\ds \mathcal{N}/Ek$ (right, see text).}  
\label{fig1}
\end{figure}

For \tck{various} levels of rotation - quantified through the Ekman number - we \tck{therefore} study how varying two parameters - the magnetic and thermal forcing terms, $\ds \widetilde{\Lambda}$ and $\ds Ra^Q$ - affects the evolution of the heat transport in the ocean, as well as its velocity. These two output quantities are tracked through their dimensionless forms, the Nusselt and Reynolds numbers. Their evolutions \tck{in our numerical model are reported} in Figures \ref{fig1a} and \ref{fig1b} for two levels of rotation. Various sets of simulations at different values of $\ds \widetilde{\Lambda}$ are considered - that is at different amplitudes of the jet for fixed rotation.
\tck{According to \citet{Gissinger2019}, temporal oscillations of the Jovian magnetic field can generate an equatorial zonal flow, whose velocity evolves as $\ds u_\phi \propto B_0^{2/3}$ or $\ds B_0^2$, depending on how numerical results are extrapolated to real planetary conditions, which in turn requires assumptions about the turbulent viscosity. This leads to magnetically-driven jet velocities to range between $\ds 1$ and $\ds 100 \, cm/s$ in Europa, \tck{comparable to estimates coming from non-magnetic convective studies \citep{Soderlund2014}}. By varying $\ds \widetilde{\Lambda}$ across unrealistically large values compared to those expected for Europa’s ocean, we can span a broader range of flow velocities without explicitly modelling turbulent viscosities, \tck{keeping in mind that the viscosity is here prescribed as a fixed (non-turbulent) value.} In order to better understand the coupling between this jet and thermal convection, we progressively increase the intensity of thermal convection with the values of the Rayleigh number.}\\

\tck{When the effects of the magnetic field are ignored ($\ds \widetilde{\Lambda}=0$), we retrieve usual results}, $\ds Nu-1$ and $\ds Re^{NA}_{polo}$ are only non zero above a critical Rayleigh number $\ds Ra_c^{HD}$. \tck{As rotation tends to stabilise convection, the value of $\ds Ra_c^{HD}$ increases like $\ds Ek^{-4/3}$ \citep{Chandrasekhar1961}. At high supercriticality $\ds Ra^Q \gg Ra_c^{HD}$ and for fixed rotation, one expects the heat flux to become independent of the distance between the two spheres as the dynamics is solely controlled by the boundary layers close to the ocean's floor or the ice \citep{Priestley1954, Malkus1954}. In our setup of fixed flux of temperature at the bottom of the ocean, this translates into $\ds Nu$ being proportional to $\ds (Ra^Q)^{1/4}$, which is similar to the regime the system would reach if it were not rotating. Likewise, boundary layer theory \tck{predicts} $\ds Re^{NA}_{polo} \propto (Ra^Q)^{1/2}$ \citep{Aubert2001}, regime which our simulations reach for $\ds Ra^Q \gg Ra_c^{HD}$.}

In the presence of the magnetically-driven jet, \tck{$\ds Nu-1$ and $\ds Re^{NA}_{polo}$} become non zero even below the threshold of convection. Indeed, the zonal jet is associated to meridional recirculation, \tck{and thus indirectly} leads to a vertical heat transport, the latter increasing with the \tck{velocity} of the flow \tck{(see Figures \ref{fig1a} and \ref{fig1b})}. \tck{We more precisely quantify the sole effect of the magnetic field on heat transport and velocity of the flow through the insets of Figure \ref{fig1a} and \ref{fig1b}, for which we only consider simulations below the threshold of convection $\ds Ra^Q = 0.03 Ra_c^{HD}$, taking respectively $\ds Ra_c^{HD}=2.10^7$ and $\ds 4.10^8$ for $\ds Ek=10^{-4}$ and $\ds 10^{-5}$. In Figure \ref{fig1a} we show that the Nusselt number only depends on the Péclet number of the flow $\ds Pe:= Pr Re^{NA}_{polo}$, with $\ds Nu-1 \propto Pe^2$ for $\ds Pe \le 100$ and $\ds Nu \propto \sqrt{Pe}$ at higher $\ds Pe$. 
\tck{As long as heat transport does not back-react on the magnetically-driven flow, the temperature field can be treated as a passive scalar, and the Nusselt number depends solely on the Péclet number. At low Péclet number, the $\ds \mb{u} \rightarrow -\mb{u}$ symmetry imposes the scaling $\ds Nu-1 \propto Pe^2$ \citep{Moffatt1983}.} The change of slope for higher $\ds Pe$ comes from the balance between the advective and diffusive terms in the temperature equation (\ref{MTH_eq_THETA_adim1}), where the diffusion only acts through boundary layers. The parallel with Europa's ocean is immediate: the zonal flow induced by electromagnetic processes is responsible for a passive heat transport through the induced meridional recirculation. We show in the inset of Figure \ref{fig1b} how this poloidal component of the velocity, characterized by $\ds Re^{NA}_{polo}$ evolves with the parameters of the problem.} \tck{The observed dependency $\ds Re^{NA}_{polo} = 0.01 \mathcal{N}/Ek$ unsurprisingly follows the prediction of \citet{Gissinger2019} $\ds u_\phi/c = 0.2 \mathcal{N}$, where $\ds \mathcal{N}:= B_0^2 \eta /(\rho \mu_0 \nu c^2)$ is the parameter they introduced and $\ds c = r_o \omega$ the phase velocity of the Jovian field in the frame of Europa}. 
\tck{This indeed can be translated into $\ds Re^{NA}_{polo} \approx (u^{NA}_{polo}/u_\phi) \mathcal{N}/Ek$, suggesting that the poloidal part of the kinetic energy, involved in the heat transport, follows the behaviour of the zonal flow, despite being significantly smaller.}

The consequences of this transport for Europa is crucial. As the electromagnetic phenomenon described here does not depend on the nature of the buoyancy, it provides an efficient mechanism for heat transport in the stably stratified layer close to the ice. Ensuring a net vertical transport in this non-convective region of the ocean may serve as a source for subsurface/surface exchanges as warm plumes would be brought to the surface to enhance melting dynamics. We emphasise that, contrary to for instance double diffusion dynamics which may also provide the ocean with vertical transport in the stably stratified layer, the magnetic pumping described here is not an instability and does not depend on a distance to some threshold to develop \citep{Gissinger2019}. This makes it a very efficient mechanism to bring warmer plumes from the ocean to the ice, \tck{as the range of $\widetilde{\Lambda}$ spanned in this study reaches expected values for Europa (Table \ref{table_method}).}

\begin{figure}[H]
\begin{center}
    \includegraphics[width=12cm]{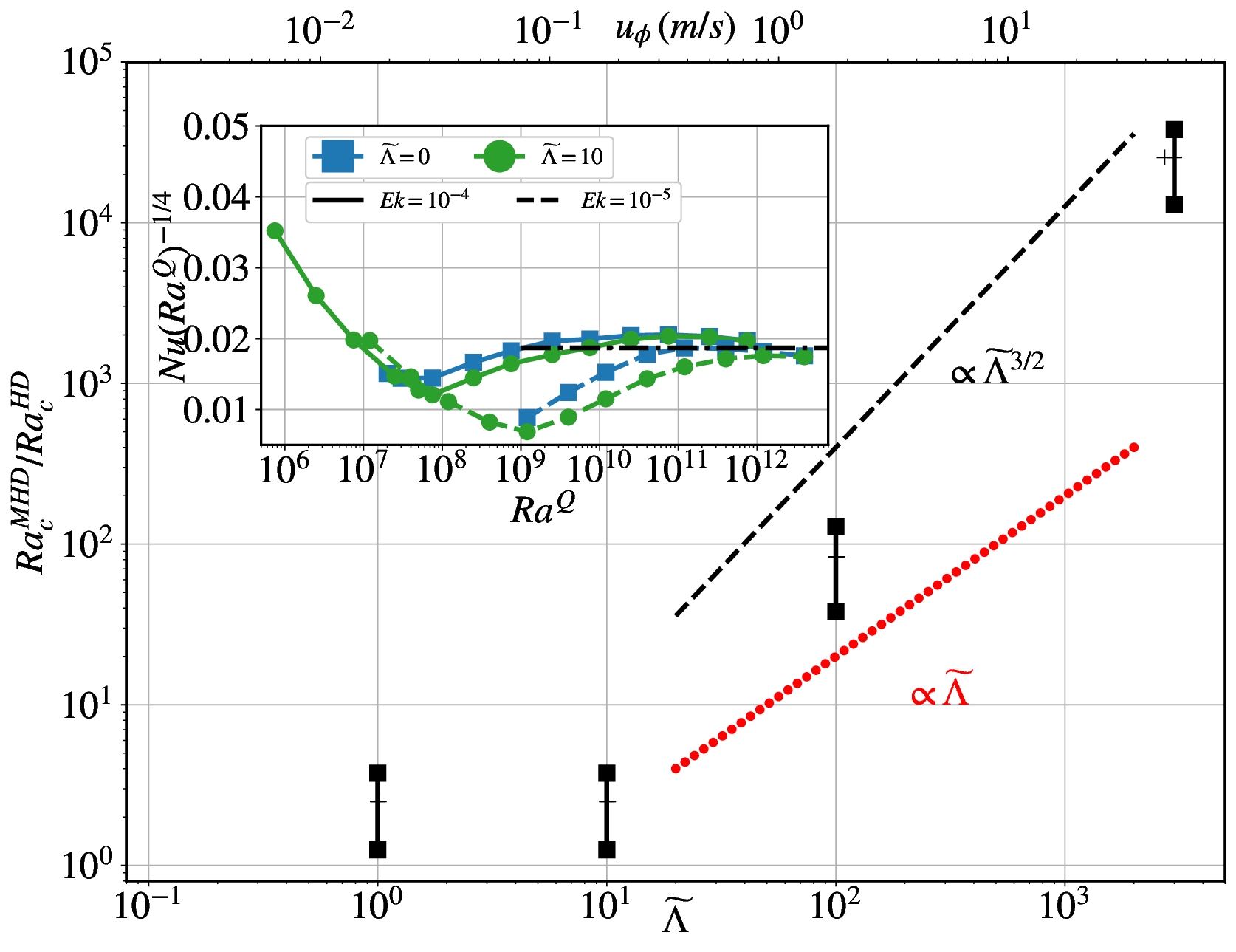}
	\caption{Modification of rotating convection thresholds. \textit{Main}: modification of the threshold of convection as a function of $\ds \widetilde{\Lambda}$ or $\ds u_\phi$ the velocity of the jet for $\ds Ek=10^{-4}$. \textit{Inset}: Compensated plot of $\ds Nu(Ra^Q)^{-1/4}$ as a function of $\ds Ra^Q$.}  
	\label{fig2a}
\end{center}
\end{figure}

In the convective region, the picture is more complex, as increasing $\ds Ra^Q$ leads to different behaviours depending on the amplitude of the zonal jet. If the moderate case of $\ds \widetilde{\Lambda}=1$ shows almost no dissimilarity with the purely hydrodynamical run, the differences increase with $\ds \widetilde{\Lambda}$. Indeed, one can see on Figure \ref{fig1a} that the presence of the magnetically-driven jet delays the value of $\ds Ra^Q$ above which the MHD case \tck{starts to follow the scaling $\ds Nu \propto (Ra^Q)^{1/4}$}. We attribute this to a modification of the onset of convection which is measured in Figure \ref{fig2a}. \tck{It is measured by reporting the value of $\ds Ra^Q$ at which $\ds Re_{polo}^{NA}$ starts to depend on $\ds Ra^Q$, namely when the curve at a given $\ds \widetilde{\Lambda}$ starts to increase.} It is tempting to interpret these results as \tck{a classical magnetoconvection effect of Jupiter's field on Europa}. Indeed, it is well known that an external magnetic field can alter the onset of convection in electrically conducting fluids. However, as shown in the \tck{main part} of Figure \ref{fig2a}, this is not what happens here. The modification of the onset \tck{we measure numerically is close to} $\ds Ra_c^{MHD} \propto \widetilde{\Lambda}^{3/2}$, a scaling that is significantly steeper than the typical magnetoconvection prediction $\ds Ra_c^{MHD} \propto \widetilde{\Lambda}$ \citep{Chandrasekhar1961, Eltayeb1972} \tck{for fixed rotation and conductivity. This is not surprising as the estimated value of the Chandrashekar number $\ds B_0^2d^2/(\rho\mu_0 \eta \nu) \approx 10$ for Europa's ocean is too small to expect magnetoconvection effects}. In other words, it is rather the presence of the jet itself that modifies the onset of convection, similar to other systems where a base flow is superimposed to thermal convection \citep{Gage1968, Bordja2010}. \tck{Even if this change is expected to be small at the value of $\ds \widetilde{\Lambda}\sim 10$ for the ocean, the uncertainty on the jet velocity might lead to more extreme changes of the threshold. In this regard, we report in the second $x$ axis of Figure \ref{fig2a} the corresponding value of the jet velocity measured in our simulations, for which we take the radial maximum of the spherically averaged $\ds u_\phi$ at the equator. The fact that all points superimpose illustrates once more the agreement with \citet{Gissinger2019}'s prediction $\ds u_\phi/c \propto \mathcal{N}$. \tck{This conversion of $\ds \widetilde{\Lambda}$ into $\ds u_\phi$ suggests that the estimate of the zonal velocity (about $\ds 1$ to $\ds 100\,cm/s$) could shift the convective threshold by one to two orders of magnitude, placing \tck{$\ds Ra_c^{MHD}$} between 1 and 100 times $\ds Ra_c^{HD}$.} } 

\tck{The presence of the jet influences not only the onset of convection but also strongly affects how the flow transitions to the non-rotating regime.} To measure at which Rayleigh number the non-rotating (NR) regime is reached, we plot in the inset of Figure \ref{fig2a} the compensated expected scaling law for two Ekman numbers, with or without magnetic fields, displaying only the value of $\ds \widetilde{\Lambda}=10$. What appears is that, \tck{consequently} to the modification of the onset of convection described above, the presence of the magnetically-driven jet delays the transition to the NR regime. In the hydrodynamical \tck{case}, this transition occurs when the system has "forgotten" about rotation and the associated preferred vertical axis that the flow tends to align with; this corresponds to Proudman-Taylor theorem \citep{Proudman1916, Taylor1917}. Here, the magnetically-driven jet somehow adds a second favoured direction - the zonal one - which \tck{opposes the radial heat flux} to modify this transition. The same behaviour is observed at a lower Ekman number.\\

\begin{figure}[H]
\begin{center}
    \includegraphics[width=12cm]{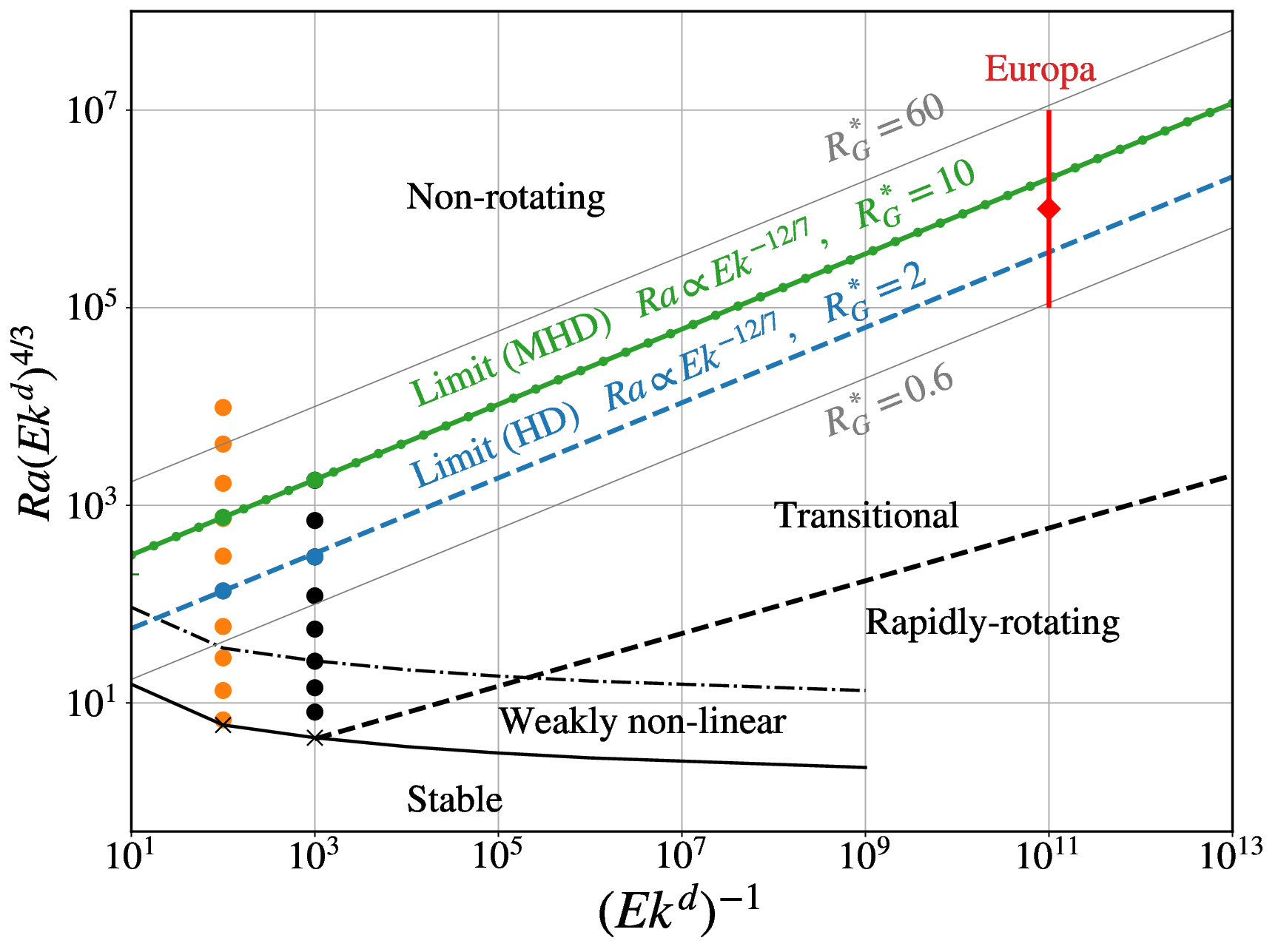}
	\caption{Regime diagram constructed with our set of simulations and inspired from \cite{Gastine2016,Lemasquerier2023,Cabanes2024}. The boundary between the NR and transitional regime is drawn using our data (see text). The lower boundary between the rapidly rotating and weakly non-linear one is taken from \cite{Gastine2016} and left as a guideline ($\ds \chi=0.6$). Possible values for the cooling pattern of Europa are reported ($\ds R_G^*$, \tck{see \ref{section_res2}}), using the estimated location of the ocean in the parameter regime given by \cite{Lemasquerier2023}.}  
	\label{fig2b}
\end{center}
\end{figure}

\tck{As shown in Figure \ref{fig2b}, the modification of the transition to the NR regime by the magnetically-driven jet has important consequences for Europa.} The diagram reported here is similar to those presented in \citet{Gastine2016, Lemasquerier2023, Cabanes2024} but it has been adapted to model Europa's ocean; \tck{besides taking into account the Jovian magnetic field, realistic no-slip and fixed-flux/temperature boundary conditions are considered, as well as an aspect ratio of $\ds \chi=0.9$ and a value of $\ds Pr=12$, expected for the ocean. The gravity profile is proportional to $\ds 1/r^2$ to model the fact that most of the mass of Europa is in its core}. \tck{The different dynamical regimes of rotating convection are reported in this diagram}.

\paragraph*{Non-magnetic case}

Close to the onset of convection, the system is said to be in the rapidly-rotating regime (RR), and its Nusselt number is notably expected to follow $\ds Nu \propto (Ra^Q)^{3/5}Ek^{4/5}Pr^{-1/5}$ \citep{Stevenson1979, Julien2012a, Bouillaut2021}. Increasing $\ds Ra^Q$ at a fixed $\ds Ek$ number eventually leads to the NR regime described earlier. The transition from RR to  NR is continuous and leads to the existence of a transitional zone in the diagram. One can draw the blue continuous boundary line by equating the two different scaling laws for each regime \citep{Gilman1977}. \tck{Our simulations reaching the NR are highlighted by blue dots and serve as a way to compute the prefactor.} Note that although this is beyond the scope of our work, the same can be done to draw the line between a weakly non-linear regime close to the onset and the RR one. We report here as a guideline the curve obtained by \citet{Gastine2016} for $\ds \chi=0.6$ and fixed temperatures that should quantitatively be modified in our setup.\\
It is possible to locate the ocean of Europa in the aforementioned diagram by evaluating the parameters of the water \tck{such as its kinematic viscosity or thermal diffusivity}, and the estimates of \citet{Lemasquerier2023} are reported in red (see also \citet{Soderlund2014, Soderlund2019} \tck{and Table \ref{table_method}}). \tck{The mean value and errorbars} were built assuming two bounds for values of the heat flux at the bottom of the ocean $\ds \Phi_i$, ranging between $\ds 6$ and $\ds 46\,mW/m^2$ (see Introduction). These situate Europa \tck{at the boundary} between the NR and transitional regime.

\paragraph*{Magnetic case}

The previous diagram \tck{is} however modified in the presence of the jet. Using the results presented in Figure \ref{fig2a}, namely that the NR regime is reached at a higher $\ds Ra^Q$ when the jet is present, the MHD green line of Figure \ref{fig2b} is built and is consequently above the blue HD one, \tck{as the prefactor is modified in the magnetised case (inset of Figure \ref{fig2a})}. This translates into Europa being pushed \tck{one order of magnitude} downwards in the diagram, away from the NR regime. As for fixed rotation $\ds Nu$ is an increasing function of $\ds Ra^Q$, this leads to \tck{one of the key findings} of this paper: the presence of the magnetically-driven jet tends to decrease the heat transfer within the ocean. \tck{By opposing the radial heat flux, its zonal direction acts as to increase the effects of rotation, somehow preventing the system from reaching the NR regime.} \tck{Note that this phenomenon is observed in our simulations corresponding to the lower estimate of the jet velocity for Europa's ocean ($\ds \widetilde{\Lambda}=10, \, u_\phi \approx 10\,cm/s$, see inset of Figure \ref{fig2a}). More extreme jet velocities may significantly amplify this decrease of the heat transfer.}

\subsection{Towards a comparison with observations}
\label{section_res2}

\paragraph*{Heat flux and ice thickness profiles}

\begin{figure}[H]
\begin{center}
    \includegraphics[width=15.5cm]{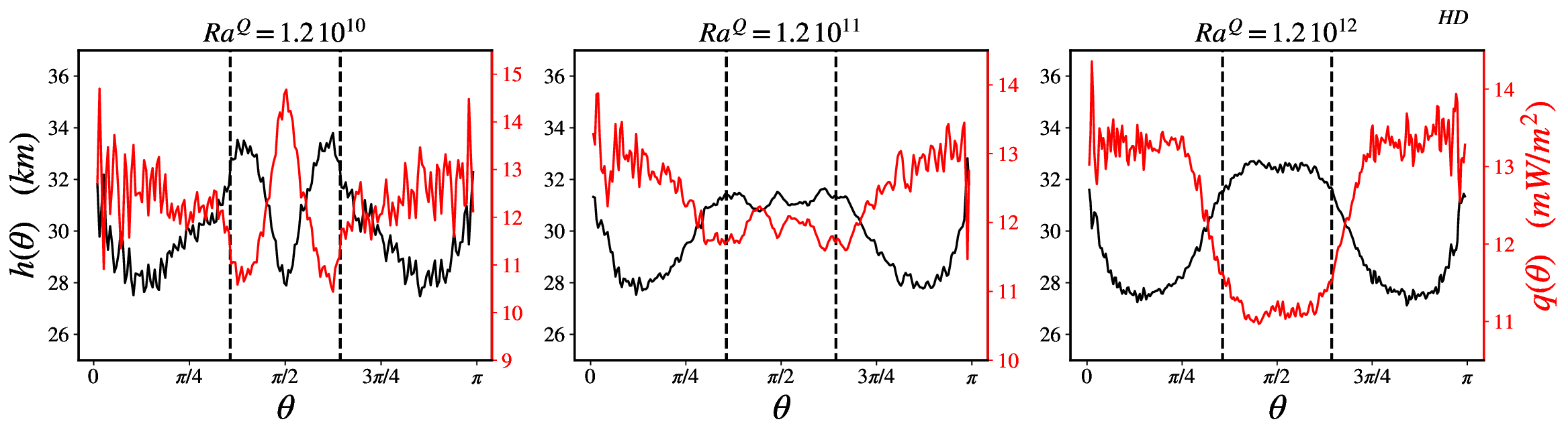} \\
    \includegraphics[width=15.5cm]{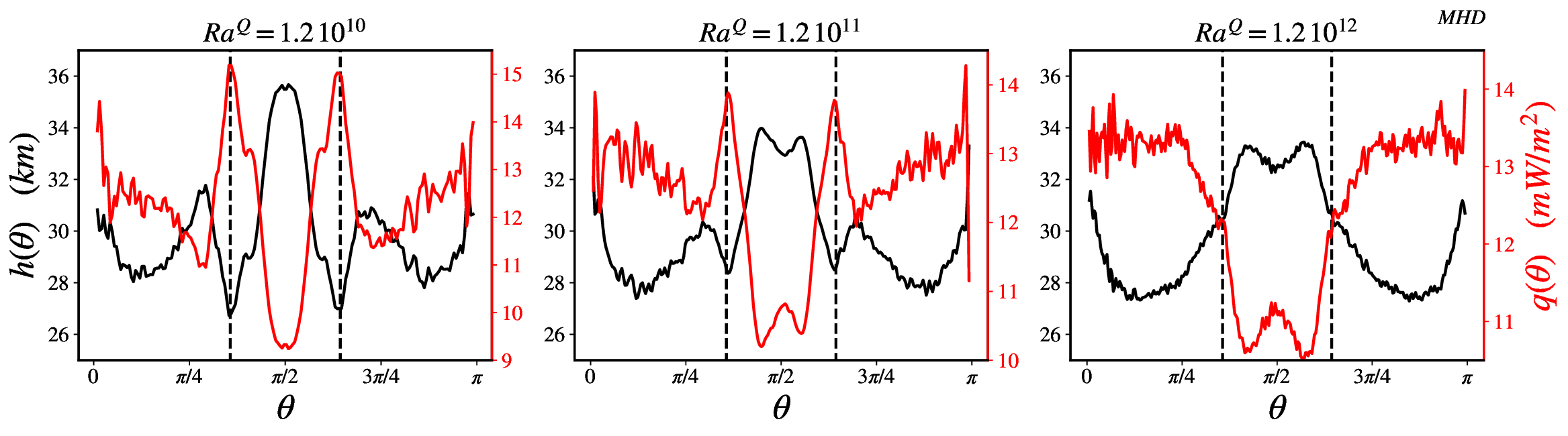}
	\caption{$\ds q=-k\partial T /\partial r$ at the ice-ocean interface (red) and corresponding thickness profiles for $\ds Ek=10^{-5}$, $\ds \widetilde{\Lambda}=10$ for $\ds \Phi_i=15\,mW/m^2$, $\ds \eta_B=10^{15}\,Pa.s$ (see Appendix \ref{appendix_ice_models}) and increasing values of $\ds Ra^Q$ from left to right, corresponding to $\ds R_G^*=\{0.2,0.7,4\}$ respectively. \tck{First and second rows correspond to purely hydrodynamical or magnetically-driven simulations respectively}. The tangent cylinder $\ds \theta_{TC}=\arcsin \chi$ and $\ds \pi-\theta_{TC}$ is emphasised with vertical dashed lines.}  
	\label{fig3}
\end{center}
\end{figure}

This global reduction of the efficiency of the transport of heat \tck{adds up to the induced ohmic heating mentioned by \citet{Gissinger2019} to illustrate the importance of electromagnetic processes in the description of Europa's ocean. Here, we describe how the jet }can have a local impact related to the heat flux spatial distribution at the interface between the ice and the ocean. Figure \ref{fig3} displays in red the heat flux $\ds q = -k\partial T/\partial r$ measured at \tck{this boundary} and averaged in azimuth for $\ds Ek=10^{-5}$ at three different values of supercriticality, both with or without magnetic field. 

In the \tck{non-magnetic} case, the heat flux is more important at the equator for the smallest $\ds Ra^Q$ reported, but this is inverted towards the poles when increasing the \tck{amplitude} of convection. These latitudinal variations are consistent with the pictures described by \citet{Kvorka2022}: close to the onset of convection, Taylor columns, parallel to the axis of rotation and located outside the tangent cylinder (TC), lead to a warmer equator \tck{heated by efficient thermal plumes coming from the mantle-ocean boundary. In contrast, the NR regime is characterised by vigorous fluid motions inside the TC, decreasing the heat transport at low latitudes in favour of the polar regions.} This equator-pole transition is observed when the relative importance of the two regimes \tck{is comparable}. \tck{\citet{Kvorka2022} proposed the parameter $\ds R_G^*=Ra(Ek^d)^{12/7}/Pr$ to measure the relative importance of NR and RR regimes. It is obtained in two steps; first by equating the two scaling laws for the Nusselt number in both regimes - assuming the transitions does not depend on $\ds Pr$, similarly to \citet{Gastine2016}, and then by dividing the result by $\ds Pr$, \tck{in order to account for the discrepancies in systems with different $\ds Pr$}. This equator-pole transition is generally expected for $R_G^*$ of order $\ds 1$, a value well verified in our simulations ($\ds R_G^*\approx 0.7$ for the middle panel of Figure \ref{fig3}).} Note that \citet{Kvorka2022} worked with a different aspect ratio $\ds \chi=0.8$ for which \tck{a reverse pole-equator transition} is later observed \tck{at higher $R_G^*$}, but we do not believe it to be relevant for Europa's ocean at the more realistic value of $\ds \chi=0.9$.

\begin{figure}[H]
\begin{center}
    \begin{subfigure}{0.45\textwidth}
        \includegraphics[width=8.0cm]{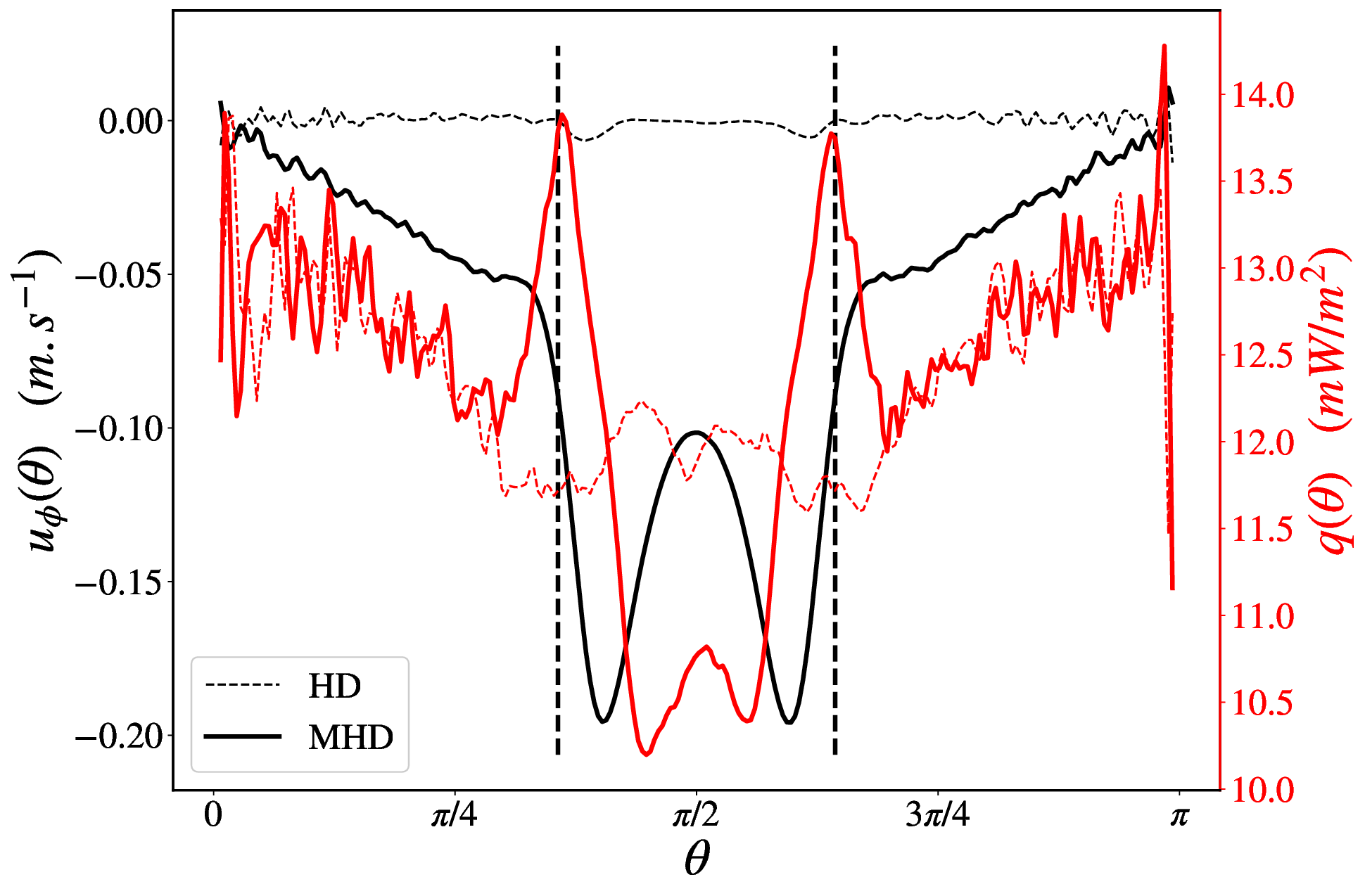}
        \caption{}
        \label{fig4a}
    \end{subfigure} 
    \begin{subfigure}{0.45\textwidth}
        \includegraphics[width=7.3cm]{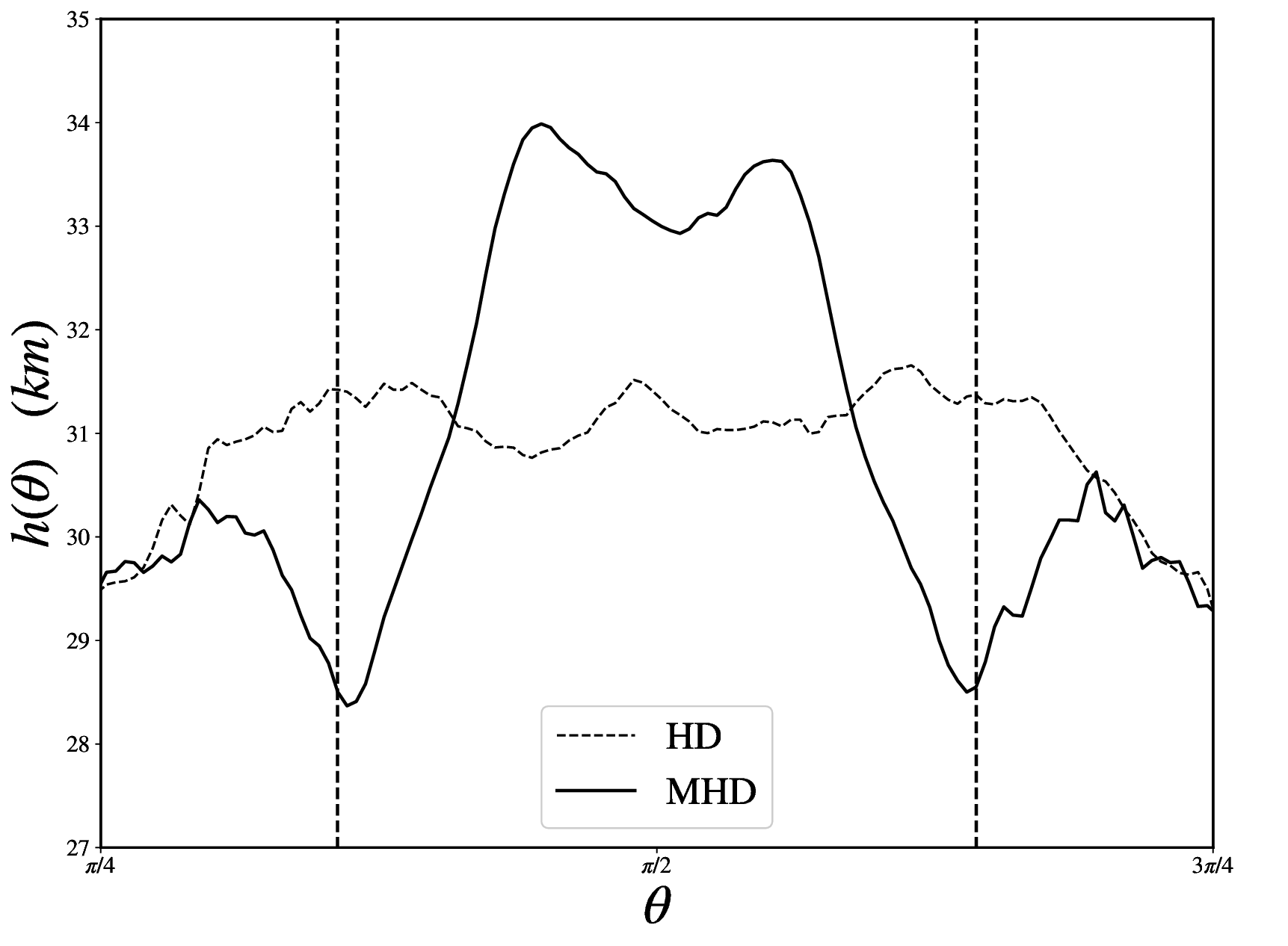}
        \caption{}
        \label{fig4b}
    \end{subfigure}
\caption{\textit{Left}: Zonal velocity at $\ds r=0.99r_o$ and heat flux at the top of the ocean for the runs of the middle panel of Figure \ref{fig3}, highlighting the anti-correlation between $\ds u_\phi$ and $\ds q$ at the TC. Due to the fact that we use no-slip boundary conditions, the same correlations between $\ds u_\phi$ and $\ds q$ cannot directly be obtained at the interface between the ocean and the ice, as was for instance done in past studies \cite{Aurnou2007, Soderlund2014} or in anelastic \tck{simulations} \cite{Raynaud2018}. The same typical peak is nevertheless observed for $\ds \theta \sim \theta_{TC}$ and $\ds \theta\sim \pi-\theta_{TC}$. \textit{Right}: Magnified comparison for the same runs of the ice thickness profile, showing the larger latitudinal variations in the MHD case.}  
\end{center}
\label{fig1}
\end{figure}

\tck{The horizontal heat flux profile is quite different when Jupiter's magnetic field is taken into account, particularly away from the NR regime:} the equatorial region consistently exhibits the lowest heat flux, which we attribute to the presence of the large scale magnetic retrograde jet centred on the equator (see Figure \ref{fig4a} in black). The most significant differences with the HD case occur for $\ds \theta_{TC} = \arcsin(\chi)$ and $\ds \pi-\theta_{TC}$ where local peaks are observed, much more important than in the non-magnetic case, corresponding to the position of the TC. \tck{These features have been observed in rotating convection under stress-free boundary conditions at the ice/ocean interface \citep{Aurnou2007} or no-slip conditions with $\ds R_G^*\ll 1$ \citep{Gastine2023}, both of which are known to promote strong zonal flows but differ significantly from Europa’s environment. Our results show that, when the magnetic field is taken into account, similar features can still emerge under parameter values relevant to Europa, owing to the magnetically-driven zonal flow.}
Here, the strong velocity disparities between the Coriolis dominated poles and the magnetically-forced equator could account for the observed behaviour at $\ds \theta_{TC}$, as suggested by the zonal velocity profile at $\ds r=0.99r_o$ which shows the same typical features. \tck{This picture is consistent with stress-free studies of convection showing a similar anti-correlation between the zonal velocity and the heat flux \citep{Yadav2016, Raynaud2018}: the shear leads to a decrease of the radial heat flux as it disperses convective structures \citep{Goluskin2014}}. %
\tck{Accurately determining the value of $\ds R_G^*$ for Europa’s ocean remains challenging, as estimates span nearly two orders of magnitude (from about $\ds 0.6$ to $\ds 60$). Interestingly, the MHD-driven peaks at the tangent cylinder persist across much of this range, including values below $\ds 1$. If future space missions constrain $\ds R_G^*$ to be of the order of unity, it would strengthen the arguments in favour of these features, which could in turn strongly affect the ice thickness, as discussed in the next section.}

\paragraph*{Ice thickness evolutions}

It is interesting to assess the importance of the latitudinal variations of the heat flux by considering models relating it to the ice thickness profile. \tck{Note that thermobaric effects, such as pressure-dependent melting at the ice-ocean interface, are neglected here. While such a pressure-dependent melting could locally alter the interface temperature profile \citep{Kang2023}, its impact on the large-scale convection is expected to be secondary compared to the dominant control exerted by the thermal Rayleigh number. This could be the subject of a future study.} Assuming that the ice is entirely \tck{thermally} conducting, \citet{Ojakangas1989} derived an analytical expression for its thickness, considering an equilibrium between tidal dissipation occurring inside the ice, the heat flux coming from the ocean and the solar radiation at the surface of the satellite. They assumed \tck{a} homogeneous \tck{heat flux} but their formula was later employed with numerical prescriptions coming from DNS outputs (e.g. \citet{Tobie2003, Lemasquerier2023}). Here we adopt a very similar approach introduced later by \citet{Nimmo2007} (see Appendix \ref{appendix_ice_models} for a brief description of Ojaganka and Stevenson's model and Nimmo's.). The ice model considered is very simple and neglects many phenomena such as \tck{changes of state} or the possibility of convection to occur inside this layer, but allows to include the effect of induction processes on the melting of the ice. \tck{This approach was confirmed recently by \citet{Gastine2025} in global simulations in which the dynamics of both the ice and the ocean are accounted for, indicating that the above model remains a good approximation to make a link between the resulting surface topography of the ice and the heat flux coming from the ocean.} 

\begin{figure}[H]
\begin{center}
    \includegraphics[width=17cm]{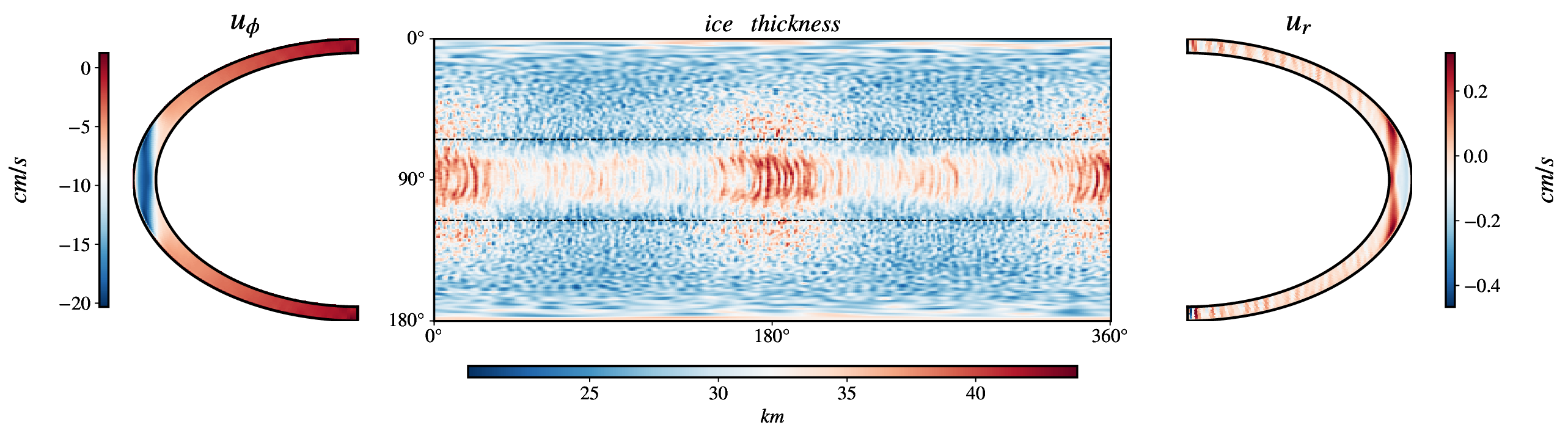}
	\caption{Geometry of the flow and resulting ice thickness profile for $\ds Ra^Q=1.2\,10^{11}$, $\ds Ek=10^{-5}$, $\ds \widetilde{\Lambda}=10$. \textit{Left (right)}: Meridional cut of $\ds u_\phi$ ($\ds u_r$) averaged in time and azimuth, showing that the equatorial region is more dynamically active that the poles. \textit{Middle}: Corresponding ice thickness profile for $\ds \Phi_i=15\,mW/m^2$ and $\ds \eta_B=10^{15}\,Pa.s$. The dominant $\ds \ell=2$, $\ds m=2$ tidal forces pattern is modified by the presence of the magnetically-driven jet as the latter induces some bulge at $\ds \theta_{TC}$ and $\ds \pi-\theta_{TC}$.}  
	\label{fig5}
\end{center}
\end{figure}

\tck{The thickness obtained from our simulations' heat fluxes are shown in Figure \ref{fig3} (black curves) and in Figure \ref{fig4b}. First, the typical reported thickness of approximately $\ds 30 \, km$ in average aligns well with current estimates for the ice layer's thickness} \citep{Pappalardo1999, Tobie2003, Soderlund2019}. Second, as the ice thickness $\ds h$ is anti-correlated to the heat flux $\ds q$, the ice shows important thickness variations at $\ds \theta_{TC}$ (Figure \ref{fig3}), \tck{which are magnified in the MHD case compared to the purely hydrodynamical one (Figure \ref{fig4b}):} \tck{the presence of the magnetically-driven jet significantly alters the distribution of ice thickness, particularly near the tangent cylinder. Figure \ref{fig5} illustrates this correlation by presenting a $\ds (\theta,\phi)$ map of the ice thickness (center), alongside meridional cuts of the velocity field (left and right).} \tck{The zonal velocity is very strong outside the TC compared to inside, whereas the radial velocity peaks at the \tck{TC} latitude, accounting for the local increase of the heat flux.} Consequently, the jet induces a variation in depth around the equator of between $\ds 5$ and $\ds 10\,km$ over only $\ds 25^o$  in latitude, which is two or three times larger than previous estimates based on \tck{non-magnetic} effects. \tck{It besides affects the general shape, which would be dominated by the $\ds \ell=2,\,m=2$ pattern of tidal dissipation occurring inside the ice} (see again Figure \ref{fig4b} and Appendix \ref{appendix_ice_models}). These strong thickness variations may influence ice fracturing in this region \citep{Greenberg1999}. \tck{As the jet is qualitatively the same below the threshold of thermal convection, the associated meridional recirculation could assure a net vertical transport in a possibly stably stratified layer close to the ice.} \\

\paragraph*{Link with topography measurements}

\begin{figure}[H]
\begin{center}
    \includegraphics[width=9.0cm]{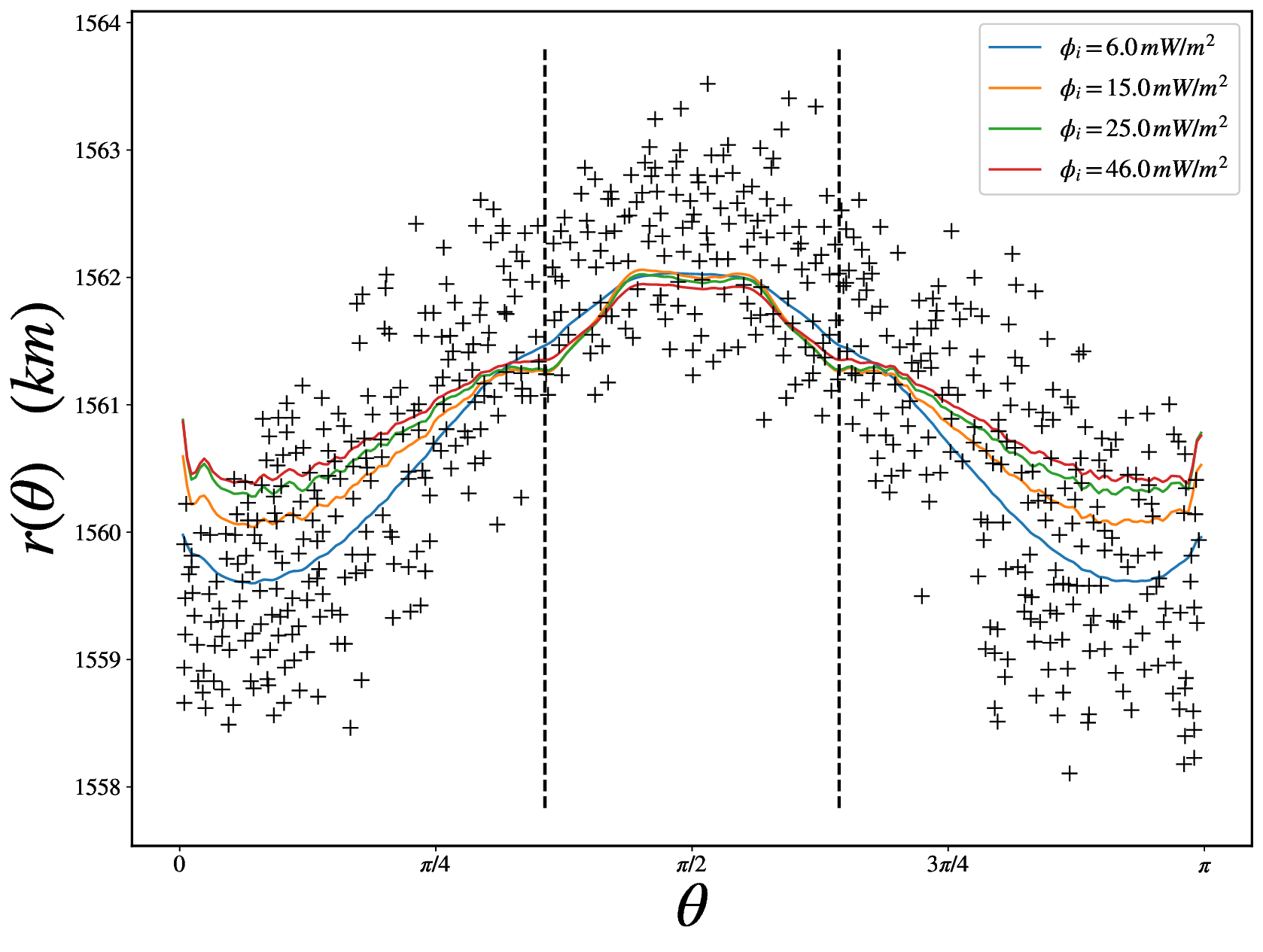}
	\caption{\tck{Azimuthally-averaged} radius profiles of Europa obtained with the simulation $\ds Ra^Q=1.2\,10^{11}$, $\ds Ek=10^{-5}$, $\ds \widetilde{\Lambda}=10$, and varying values of the heat flux imposed at the ocean's floor for $\ds \eta_B=10^{15}\,Pa.s$. A comparison is made against observations of limb profiles reported by \cite{Nimmo2007}, \tck{at one longitude only due to currently limited observations}. For the highest values of $\ds \Phi_i$, a bulge appears in the vicinity of the tangent cylinder which could serve as a probe for an indirect inferring of the ocean's geometry.}  
	\label{fig4}
\end{center}
\end{figure}

\tck{Figure \ref{fig4} compares our simulations to current observations and allows us to make several predictions.
\tck{To make the link with the observations reported, $\ds r(\theta,\phi) + \Delta \rho/(\rho + \Delta \rho) \left(h-h_{ref} \right)$ is plotted, where $\ds \rho$ is the ice density and $\ds \Delta \rho$ is the difference between the ice and water density. $\ds h_{ref}$ is a reference height. $\ds r(\theta,\phi)$ is the general shape of an ellipsoid obtained by \citet{Nimmo2007} for Europa, whose description is re-given in Appendix \ref{appendix_Nimmo_obs}}. \tck{Following earlier discussions,} the simulation we consider is estimated to be possibly relevant for Europa's ocean, as it reproduces its values of $\ds Ek_\eta$, $\ds \widetilde{\Lambda}$ and $R_G^*$. First, we find that our model shows overall good agreement with existing data on Europa’s topography, as inferred from ellipsoidal fits \citep{Nimmo2007}. However, it begins to diverge when the heat flux $\Phi_i$ at the base of the ocean exceeds $45\, mW/m^{2}$ (red line). This trend is consistent with the analytical work of \citet{Ojakangas1989}, who demonstrated that ice thickness results from a balance among solar insolation, oceanic heat flux, and tidal dissipation in the ice. Their theory provides a critical flux estimate of order $10\,mW/m^{2}$, above which tidal dissipation smooths out \tck{other} horizontal thickness variations. Consequently, our simulations suggest an upper limit on the oceanic heat flux for a given ice viscosity. Second, our results predict the emergence of a bulge around the tangent cylinder in the ocean - an effect absent in \tck{non-magnetic} models of Europa's ocean. Although current measurements \citep{Davies1998, Nimmo2007} cannot definitively confirm or rule out this bulge, upcoming missions such as JUICE and Europa Clipper may achieve the resolution needed to detect it. In that case, the presence of a bulge would provide an indirect measurement of the ocean geometry, specifically the ratio $\chi=r_i/r_o$; for a typical ice thickness of $\ds 30 \,km$ and an aspect ratio varying between $\ds 0.9$ and $\ds 0.94$, it would lead to an ocean's depth between $\ds 90$ to $\ds 150\,km$.  Conversely, if no bulge is observed, it would constrain the parameter $R_G^*$ for Europa's ocean, as high $R_G^*$ values prevent bulge formation in our simulations.}

\section{Discussion}
\label{section_discussion}

In the perspective of the upcoming JUICE and Europa Clipper missions, gaining insight into the ocean’s influence on Europa’s ice shell is essential. \tck{Modelling the combined effects of the Jovian magnetic field and thermal convection on Europa's ocean, we showed that the threshold of convection in this environment is modified compared to non-magnetic studies. Besides, the efficiency of the heat transport due to the magnetic field proves to be critical in stably stratified layers of the ocean to bring warm plumes to the ice. We studied the evolution of the heat flux profile due to the non-linear coupling between convection and electromagnetic processes. Showing that magnetic cases tend to be more rotationally constrained that their purely hydrodynamical counterparts, we then used idealised models relating the heat flux at the top ocean to the ice thickness evolution.}
In the magnetised case, our results show more pronounced latitudinal variations in ice thickness, centred on the latitude of the tangent cylinder. These could possibly favour very localised ice fracturing at the surface.
If one were able to obtain a map of the thickness of the ice, or more reasonably a global map of the topography of the satellite, as was for instance done for Titan \citep{Corlies2017, Durante2019}, one could therefore relate these observations to the dynamics of the subsurface ocean and infer some of its internal properties.  \\
Our results must however be taken with caution: the ice-ocean interaction could change the general dynamics. Besides, the implications on the thickness of the superficial layer assume an ice being entirely conducting, while convection inside the ice would smoothen thickness variations as the melting ice is replenished by convection of the ice \citep{Kvorka2022, Kihoulou2023}. Regarding the dynamics of the ocean itself, tides are obviously a missing ingredient of our description \citep{Tyler2008, Lemasquerier2023}, as well as libration, inertial waves, or double diffusion processes. We are currently developing a unified numerical model including convection, rotation, magnetic fields, and tidal forcing.\\
Nevertheless it is important to point out that our results may apply to a wider class of problems dealing with the interaction of zonal flows and convection. The magnetically-driven jet for instance shows some similarities with tidal flows, being both zonal and with the same $\ds m=2$ azimuthal symmetry \citep{Gissinger2019}, as well as function of the nearby planet spin (see for instance \citet{Astoul2022}). \tck{It could for instance apply to other Jovian moons such as Ganymede or Callisto, or even other icy satellites for which the planet's magnetic field is non-axisymmetric. See the discussion in \citet{Soderlund2024}}. 
\tck{In addition, as the magnitude of the magnetically-driven jet remains poorly constrained, yet appears capable of inducing significant effects, this study highlights the need to incorporate MHD effects into models of Europa’s subsurface ocean. \citet{Hay2023} indeed related zonal velocity inside the ocean - but originating from thermal convection - with torques applied on the ice that \tck{could have} impacted non-synchronous rotation \tck{throughout the history of the satellite}. \tck{An important perspective for future work concerns the enhanced topographic variations observed in the MHD regime. If this topography were explicitly incorporated into the model, as in \citet{Gastine2025}, it could lead to amplified hydrodynamic feedbacks and potentially modify the large-scale flow dynamics.} Global 3D DNS taking into account convection, MHD as well as melting dynamics therefore seems to be the next promising step in order to understand future measurements.}

\section*{Acknowledgements}
This work was supported by funding from the French program
'JCJC' managed by Agence Nationale de la Recherche (Grant No. ANR 19-CE30-0025-01), from CEFLPRA contract 6104-1 and the Institut Universitaire de France. This study used the HPC resources of MesoPSL financed by the Région île-de-France and the project EquipMeso (reference ANR-10-EQPX-29-01) of the program Investissements d'Avenir, supervised by the Agence Nationale pour la Recherche. We thank two anonymous referees for their comments that led to improve the quality of this paper.

\appendix
\counterwithout{equation}{section}
\setcounter{equation}{0}

\section{Jupiter's magnetic field $\ds \mb{B_0(t)}$}
\label{appendix_B0}

\tck{We recall here for completeness the expression of the magnetic field of Jupiter in the reference frame of Europa. We refer to \citet{Gissinger2019} for the original idea. In its dimensionless form it reads:}


\begin{align}
B_{0r} & =\frac{2 \sin \theta \cos (\omega t-\phi)}{r_j^3}-\frac{3 \sin \theta \cos \phi \cos \omega t}{r_j^5} +\frac{3 \gamma r}{r_j^5}\left[\sin \omega t-\sin ^2 \theta \sin \phi \cos (\omega t-\phi)\right], \label{ch_EURO_DNS_eq_B0r}\\    
B_{0\theta} & =\frac{2 \cos \theta \cos (\omega t-\phi)}{r_j^3}-\frac{3 \cos \theta \cos \phi \cos \omega t}{r_j^5} -\frac{3 \gamma r}{r_j^5}[\sin \theta \cos \theta \sin \phi \cos (\omega t-\phi)+ \gamma r \cos \theta \cos (\omega t-\phi)], \label{ch_EURO_DNS_eq_B0t}\\
B_{0\phi} & =\frac{2 \sin (\omega t-\phi)}{r_j^3}+\frac{3 \sin \phi \cos \omega t}{r_j^5} +\frac{3\gamma r }{r_j^5}[\sin \theta \cos \omega t-\sin \theta \sin \phi \sin (\omega t-\phi)- \gamma r \sin (\omega t-\phi)] \label{ch_EURO_DNS_eq_B0p},
\end{align}

having introduced $\ds r_j = \left(1 + \gamma^2 r^2+2\gamma r\sin \theta \sin \phi \right)^{1/2}$. $\gamma=2.10^{-3}$ is the ratio between Europa's radius and the distance Jupiter-Europa. \\

\section{Ice thickness models}
\label{appendix_ice_models}
\renewcommand{\thesubsection}{\thesection.\arabic{subsection}}

\subsection{Nimmo's model}
\label{appendix_Nimmo}

We recall here the ideas of \citet{Nimmo2007}'s model for completeness. The reader is referred to the original publication for more details. A diffusion equation is considered for the  \tck{temperature of the }ice of the form :

\begin{equation}
	\frac{\partial}{\partial z} \left(k_{ice} \frac{\partial T}{\partial z} \right)= -Q,
	\label{method_eq_model1}
\end{equation}

where $\ds z$ is the vertical direction. $\ds k_{ice} = 3\, W.m^{-1}.K^{-1}$ is the thermal conductivity of the ice. As \citet{Nimmo2007} , we do not take into account the variation of $\ds k_{ice}$ with temperature as they showed that it would lead to comparable results. $\ds Q$ is the tidal volumic dissipation rate. Its expression is :
\begin{equation}
	Q(\theta,\phi) = \frac{2 \mu \overline{\dot{\varepsilon_{ij}}^2}(\theta,\phi)}{\omega}\left(\frac{\omega \tau_M}{1 + (\omega \tau_M)^2} \right),
	\label{method_eq_model2}
\end{equation}

with $\ds \mu = 4\,GPa$, $\ds \omega=2.05\,rad.s^{-1}$ the tidal frequency. $\ds \tau_M = (\eta_B/\mu) \exp \left(-\gamma(T-T_o) \right)$ accounts for the variation of the viscosity of the ice with temperature, with $\ds \gamma= 0.1\, K^{-1}$ and $\ds T_o=270\,K$ the temperature at the base of the ice. $\ds \eta_B$ ranges between $\ds 10^{14}$ and $\ds 10^{15}\,Pa.s$. $\ds \dot{\varepsilon_{ij}}$ is the local strain rate tensor for a Maxwell medium, expressed in the Appendix of \citet{Ojakangas1989}, \tck{the $\dot{\tcw{X}}$ corresponding to a time derivative and the $\overline{\tcw{X}}$ to a time average, following the original notations}. It leads the quantity $\ds Q$ to correspond to a $\ds \ell=2,\,m=2$ pattern.\\
The ice thickness profile is then obtained by integrating Eq.(\ref{method_eq_model1}) for a semi-infinite domain vertically, imposing $\ds T=T_o$ and $\ds k\partial T/\partial z = q$ at $\ds z=z_{ice}$, with $\ds q$ the heat flux taken from a simulation. With our choice of dimensionless units, it is expressed as $\ds q= - \Phi_i \chi^2 \partial \widetilde{T}/\partial \widetilde{r}(r=r_o)$, where $\ds \widetilde{..}$ denotes an output from a simulation. It varies both in latitudes and longitudes (see Figure \ref{fig5}). The height at which the temperature matches that of the surface is the local thickness. The surface temperature $\ds T_S$ corresponds to a time average and is obtained by equilibrating the part of the solar flux which is absorbed by the ice with the heat flux which is radiated away, taking besides into account the obliquity of the satellite. It leads to an average surface temperature profile which is symmetric with respect to the equator and varies between $\ds \sim 50 K$ at the poles and $\ds \sim 100K$ at the equator \citep{Ojakangas1989}.

\renewcommand{\thesubsection}{B.\arabic{subsection}}
\subsection{Ojakangas and Stevenson's model}
\label{appendix_OS}

Considering Eq.(\ref{method_eq_model1}), \citet{Ojakangas1989} were able to obtain an analytical expression for the ice thickness profile $\ds h$, considering some simplification for the viscosity of the ice, expressed as a power law of temperature. With the same notations it reads:

\begin{equation}
	h(\theta,\phi) = \frac{ln(T_o/T_S)}{\sqrt{\left(\frac{2}{a_1}\right)\int_0^{T_o} \frac{Q(T)dT}{T} + \left(\frac{q}{a_1}\right)^2 }},
	\label{eq_method_h}
\end{equation}

where $\ds a_1 = kT = 4.88\,10^2\,kg.m.s^{-3}$ as they consider the variations of the conductivity with temperature. The orders of magnitude discussed in the main text is obtained by comparing the contribution from each of the terms at the denominator. 




\refstepcounter{section}
\renewcommand{\thesection}{Appendix \Alph{section}}

\begin{center}
  \textbf{\thesection: Topography and comparisons with observations}
\end{center}

\addcontentsline{toc}{section}{Appendix \Alph{section}: Topography and comparisons with observations}
\label{appendix_Nimmo_obs}
\vspace{1ex}

The general shape of Europa being that of an ellipsoid, the radial distance of a point at its surface is given by $\ds r^2(\theta,\phi) = \sin^2 \theta \left(a^2\cos^2\phi + b^2 \sin^2\phi \right) + c^2 \cos^2 \theta$, with $\ds a,b$ and $\ds c$ the three axes of the ellipsoid and $\ds \theta$ and $\ds \phi$ the colatitude and longitude respectively. Using constraints obtained from gravity measurements \citep{Anderson1998}, the assumption that it is a synchronous rotator in hydrostatic equilibrium \citep{Murray1999} and limb observations, \citet{Nimmo2007} reported ranges of acceptable values for $\ds a,b$ and $\ds c$. In this work, we consider $\ds a=1561.6\,km$, $\ds b=1560.1\,km$ and $\ds c=1559.3\,km$. The formula $\ds r(\theta,\phi) + \Delta \rho/(\rho + \Delta \rho) \left(h-h_{ref} \right)$ which is plotted in Figure \ref{fig4} makes use of the assumption of isostasy relating the ice thickness variations to the local topography \citep{Nimmo2007}. $\ds h_{ref}$ is a reference thickness. It changes with $\ds \Phi_i$, as increasing the flux leads to a thinning of the ice. For the curves of Figure \ref{fig4}, we used $\ds h_{ref} =32.75,\,21.5,\,12.5,\,3.5\,km $ for $\ds \Phi_i=6,\,15,\,25,\,46\,mW/m^2$ respectively, to match with the observations. Note that the curves reported are averaged in the zonal direction, whereas observations correspond to a specific longitude. It means that azimuthal variations could be expected with this approach, as shown with the map of the thickness in Figure \ref{fig5}b, but these cannot be captured yet due to a too little number of observations.



\bibliographystyle{unsrtnat}
\bibliography{ref}
\end{document}